\documentclass[preprint,showpacs,preprintnumbers,prb,amsmath,amssymb]{revtex4}

\usepackage{graphicx}
\usepackage{dcolumn}
\usepackage{bm}

\begin{document}

\preprint{APS/123-QED}

\title{Electronic States in Silicon Quantum Dots:\\
Multivalley Artificial Atoms}

\author{Yoko Hada}
\email{yokoh@rk.phys.keio.ac.jp}
\author{Mikio Eto}%
\affiliation{%
Faculty of Science and Technology, Keio University,\\
3-14-1 Hiyoshi, Kohoku-ku, Yokohama 223-8522
}%

\date{\today}

\begin{abstract}
Electronic states in silicon quantum dots are examined theoretically,
taking into account a multivalley structure of the conduction band.
We find that (i) exchange interaction hardly
works between electrons in different valleys. In consequence
electrons occupy the lowest level in different valleys in the absence
of Hund's coupling when the dot size is less than 10 nm.
High-spin states are easily realized by applying a small magnetic field.
(ii) When the dot size is much larger, the electron-electron interaction
becomes relevant in determining the electronic states. Electrons are
accommodated in a valley, making the highest spin, to gain the exchange
energy. (iii) In the presence of intervalley scattering,
degenerate levels in different valleys are split. This could result in
low-spin states. These spin states in multivalley artificial atoms can
be observed by looking at the magnetic-field dependence of peak positions
in the Coulomb oscillation.
\end{abstract}

\pacs{Valid PACS appear here}
\maketitle

\section{Introduction}
Quantum dots fabricated on semiconductor heterostructures, {\rm e.g.},
GaAs/AlGaAs, are called artificial atoms.
The electronic states in such quantum dots have been elucidated by
examining the Coulomb oscillation;
a peak structure of the current as a function of gate voltage
which changes the number of electrons in the dots, $N$, one by one.
In disk-shaped quantum dots,
discrete energy levels show an atomic-like shell structure.\cite{Ta}
They are filled consecutively with increasing $N$.
When a shell is partly filled,
electrons occupy degenerate energy levels
with parallel spins (Hund's rule),
as seen in the periodic table of atoms.\cite{Ta,Kouwenhoven}
These properties of the artificial atoms have been quantitatively
explained by theoretical studies including the electron-electron Coulomb
interaction exactly.\cite{qd1,qd2,qd3,qd4,Eto1,Eto2,Eto3}

Compared with the electronic states in GaAs quantum dots,
those in silicon (Si) quantum dots have been poorly understood.
A combination of microfabrication and oxidation techniques on Si
enables to fabricate so small quantum dots
as $\sim 10$ nm.\cite{Tak,Ho,Hi,Sa,Ro1,Ro2,On}
The Si quantum dots are being extensively studied for their
application to single-electron devices working at room temperature.
However, the lack of information on the precise shape of
individual dots has
prevented fundamental researches of the electronic states.
Particularly, there have been few theoretical studies which consider
the effects of a multivalley structure of the conduction band on the
electronic states in Si quantum dots.\cite{Na}
The multivalley structure should lead to different properties of
electronic states from those in GaAs quantum dots with single valley
at $\Gamma$ point.

Recently some experiments of the Coulomb oscillation have been reported
for Si quantum dots.\cite{Ro1,Ro2,On}
The magnetic-field dependence of the current peaks indicates
various spin states in the quantum dots.
Another experiment has implied the observation of the Kondo effect,
which might be due to degenerate energy levels by the multivalley
structure of the conduction band.\cite{Ro3}
The explanation for these phenomena requires the understanding of the
electronic states in Si quantum dots.
The study of spin states is also important for the application of
the quantum dots to ``spintronics," quantum computers,\cite{Loss} etc.

The purpose of the present paper
is to elucidate fundamental characters of the electronic states
and spin configurations in Si dots, ``multivalley artificial atoms,''
for the first time.
We adopt the effective mass approximation,\cite{Lu} assuming
that the size of quantum dots is sufficiently larger than the lattice
constant, to consider the following properties of Si;
(i) more than one valley of the conduction band
(two equivalent valleys in Si-MOS systems and six in the bulk),
(ii) anisotropy of the effective mass at a valley
($m^*_l=0.98m_0$, $m^*_t=0.19m_0$), and
(iii) large Zeeman effect (effective g-factor is $g^*\simeq 2$ in Si,
whereas $g^*\simeq 0.4$ in GaAs).
We determine the electronic configuration of the ground state,
considering Coulomb and exchange interactions.
We show that the exchange interaction hardly works between electrons in
different valleys. In consequence electrons occupy the lowest level in
different valleys in the absence of Hund's coupling when the dot size
is less than 10 nm. High-spin states are easily realized by applying a small
magnetic field. This is quite different from the case of
single-valley artificial atoms of GaAs quantum dots, where high spins
appear due to the exchange interaction only in the case of orbital
degeneracy (Hund's rule).
When the dot size is much larger than 10 nm, the electron-electron interaction
becomes relevant in determining the electronic states. Electrons are
accommodated in a valley, making the highest spin, to gain the exchange
energy.

We also investigate the effects of anisotropic shape of quantum dots and
intervalley scattering. The former tends to make higher spins, whereas the
latter could make lower spins. These spin states in multivalley artificial
atoms should be observed by looking at the magnetic-field dependence of the
peak positions in the Coulomb oscillation. We calculate the addition energies,
as functions of the magnetic field, which correspond to the peak positions
in the Coulomb oscillation.

The organization of the present paper is as follows.
In the next section (Sec.\ \ref{sec:model}), we explain our model
for Si quantum dots and calculation method using the effective mass
approximation. In Sec.\ \ref{sec:size}, we present the calculated
results for the electronic states in isotropic Si quantum dots.
We discuss the fundamental properties of multivalley artificial
atoms, with changing the size of the dots. In Sec.\ \ref{sec:shape},
we examine anisotropic Si quantum dots.
In Sec.\ \ref{sec:intervalley}, effects of intervalley scattering on
the electronic states are considered.
The last section (Sec.\ \ref{sec:conclusion}) is devoted to our
conclusions and discussion.

\section{Model and calculation method}
\label{sec:model}

For the multivalley structure of conduction band in Si,
we assume that two valleys at $\bm{k}=(0,0,\pm k_0)$,
where $k_0=0.85 \times 2\pi / a$ ($a$ is the lattice constant, $0.543$ nm),
are equivalent and located below the other valleys at $\bm{k}=(\pm k_0,0,0)$
and $(0,\pm k_0,0)$ by $\Delta E_{\rm valley}$.
$\Delta E_{\rm valley}=100$ meV,
considering the effect of oxidation-induced strain.\cite{Ho,Shi}
The valleys of $\bm{k}=(0,0,\pm k_0)$, $(\pm k_0,0,0)$, and $(0,\pm k_0,0)$
are denoted by $\pm k_z$, $\pm k_x$, and $\pm k_y$ hereafter.

As a model for the confinement potential of quantum dots,
we adopt a three-dimensional harmonic potential $V(x,y,z)$.
We examine an isotropic confinement
\begin{equation}
V(x,y,z)=\frac{1}{2}K(x^2+y^2+z^2),
\label{eq:isoV}
\end{equation}
to focus on generic properties of electronic states in Si dots,
for a while. An anisotropic confinement is studied later
(Sec.\ \ref{sec:shape}).

First, we calculate one-electron states and energy levels in the absence of
electron-electron interaction.
In the effective mass approximation,\cite{Lu}
the wavefunction for electrons around a valley,
{\rm e.g.}, $\pm k_z$, is described by a product of the
Bloch wave at the valley and the envelope function $F_{\pm k_z}$
\begin{eqnarray}
\psi_{\pm k_z}(\bm{r})= F_{\pm k_z}(\bm{r})
e^{\pm i \bm{k}_z \cdot \bm{r}} u_{\pm k_z}(\bm{r}),
\end{eqnarray}
where $\bm{k}_z=(0,0,k_0)$ and
$u_{\pm k_z}(\bm{r})$ is a function with the periodicity of
the Bravais lattice.
The envelope function $F_{\pm k_z}$ satisfies the effective mass
equation
\begin{eqnarray}
&& \left[-\frac{\hbar^2}{2m_t^*}(\frac{\partial^2}{\partial x^2} + 
\frac{\partial^2}{\partial y^2})
-\frac{\hbar^2}{2m_l^*}\frac{\partial^2}{\partial z^2} + 
V(x,y,z) \right]  F_{\pm k_z}(\bm{r})
\nonumber \\
&=& \varepsilon
F_{\pm k_z}(\bm{r}),
\label{eq:EMequ}
\end{eqnarray}
where the energy $\varepsilon$
is measured from the bottom of conduction band at $\pm k_z$.
[In the effective mass equation around the valley of $\pm k_x$ ($\pm k_y$),
$m^*_l$ appears in the differential term with respect to $x$ ($y$).]
We disregard the interband scattering on the assumption that 
the confinement potential $V(x,y,z)$ is so smooth that its
Fourier components with $k \sim 1/a$ are negligibly small.
(The effect of interband scattering, which may stem from impurities
within the dots, sharp edges of the dots, etc., is discussed in
Sec.\ \ref{sec:intervalley}.)
Using $V(x,y,z)$ in Eq.\ (\ref{eq:isoV}), Eq.\ (\ref{eq:EMequ}) yields
the eigenvalues of
\begin{eqnarray}
\lefteqn{\varepsilon(\pm k_z; n_x,n_y,n_z)}
\nonumber \\
&=& \hbar \sqrt{\frac{K}{m^*_t}}(n_x +\frac{1}{2})
+ \hbar \sqrt{\frac{K}{m^*_t}}(n_y +\frac{1}{2})
+ \hbar \sqrt{\frac{K}{m^*_l}}(n_z +\frac{1}{2}),
\nonumber \\
\ 
\label{eq:level}
\end{eqnarray}
with $n_x$, $n_y$, $n_z=0,1,2,\cdots$.
[For valleys of $\pm k_x$ ($\pm k_y$), $m^*_l$ appears in the first
(second) term.] The eigenfunction for the lowest level with
$(n_x,n_y,n_z)=(0,0,0)$ is given by
\begin{eqnarray}
F_{\pm k_z}^{\, 0,0,0}(\bm{r}) &=&
(\frac{1}{\pi \hbar})^\frac{3}{4}
({m_t^*}^2 m_l^* K^3)^\frac{1}{8} \nonumber \\
&& \times \exp \left[-\frac{1}{2\hbar} ( \sqrt{m_t^* K} x^2 +
\sqrt{m_t^* K} y^2 + \sqrt{m_l^* K} z^2 ) \right].
\nonumber \\
\ 
\end{eqnarray}

The size of the confinement potential is given by
\begin{eqnarray}
l=2\sqrt{\hbar/m^*_t \omega_t},
\end{eqnarray}
where $\omega_t=\sqrt{K/m^*_t}$, using the smaller effective mass $m^*_t$.
Three cases of $l=$5 nm, 10 nm, and 15 nm are examined.

Many-body states are simply represented by single configuration
of the occupation of the one-electron levels by $N$ electrons,
for semi-quantitative discussion.
The Coulomb and exchange interactions between electrons $(k_i; n_x,n_y,n_z)$
and $(k_j; n_x',n_y',n_z')$ are written as
\begin{eqnarray}
&& \int \int d\bm{r}_1 d\bm{r}_2
|{\psi_{k_i}^{n_x,n_y,n_z}} (\bm{r}_1)|^2
\frac{e^2}{4\pi\varepsilon \mid \bm{r}_1-\bm{r}_2 \mid}
|{\psi_{k_j}^{n_x',n_y',n_z'}} (\bm{r}_2)|^2
\nonumber \\
&=& \int \int d\bm{r}_1 d\bm{r}_2
|{F_{k_i}^{n_x,n_y,n_z}}(\bm{r}_1)|^2
\frac{e^2}{4\pi\varepsilon\mid \bm{r}_1-\bm{r}_2 \mid }
|{F_{k_j}^{n_x',n_y',n_z'}}(\bm{r}_2)|^2
\nonumber \\
\ 
\label{eq:Coulomb}
\end{eqnarray}
and
\begin{eqnarray}
&& \int \int d\bm{r}_1 d\bm{r}_2
{\psi_{k_i}^{n_x,n_y,n_z}}^* (\bm{r}_1)
{\psi_{k_j}^{n_x',n_y',n_z'}}^* (\bm{r}_2)
\nonumber \\
&& \times \frac{e^2}{4\pi\varepsilon \mid \bm{r}_1-\bm{r}_2 \mid}
\psi_{k_j}^{n_x',n_y',n_z'}(\bm{r}_1)
\psi_{k_i}^{n_x,n_y,n_z}(\bm{r}_2)
\nonumber \\
&=& \int \int d\bm{r}_1 d\bm{r}_2
e^{i(\bm{k}_j-\bm{k}_i)\cdot \bm{r}_1}
e^{i(\bm{k}_i-\bm{k}_j)\cdot \bm{r}_2}
\frac{e^2}{4\pi\varepsilon\mid \bm{r}_1-\bm{r}_2 \mid }
\nonumber \\
&& \times \ {F_{k_i}^{n_x,n_y,n_z}}^*(\bm{r}_1)
{F_{k_j}^{n_x',n_y',n_z'}}^* (\bm{r}_2)
F_{k_j}^{n_x',n_y',n_z'}(\bm{r}_1)
\nonumber \\
&& \times  \ F_{k_i}^{n_x,n_y,n_z}(\bm{r}_2)
\ u_{k_i}^*(\bm{r}_1) u_{k_j}^*(\bm{r}_2)
u_{k_j}(\bm{r}_1) u_{k_i}(\bm{r}_2), 
\label{eq:exchange}
\end{eqnarray}
respectively, with dielectric constant $\varepsilon=11.9 \varepsilon_0$.
These terms are evaluated numerically. We calculate the total energy
for all possible configurations and determine the ground state.

In a magnetic field $\bm{B}=(0,0,B)$, we consider the Zeeman energy
\begin{eqnarray}
E_{\rm Zeeman} = -g^* \mu_{\rm B} S_{{\rm tot},z} B,
\label{eq:Zeeman}
\end{eqnarray}
where $g^*=2$, $\mu_{\rm B}$ is the Bohr magneton, and $S_{{\rm tot},z}$ is the
$z$ component of the total spin.
The orbital magnetization effect is disregarded
because of large effective mass in Si.\cite{orbital}

\section{Electronic states in Si quantum dots}
\label{sec:size}

In this section, we clarify the basic properties of electronic states
in Si quantum dots, considering an isotropic confinement, Eq.\ (\ref{eq:isoV}).

\subsection{Electronic states}

\begin{figure}
\includegraphics[scale=1.00]{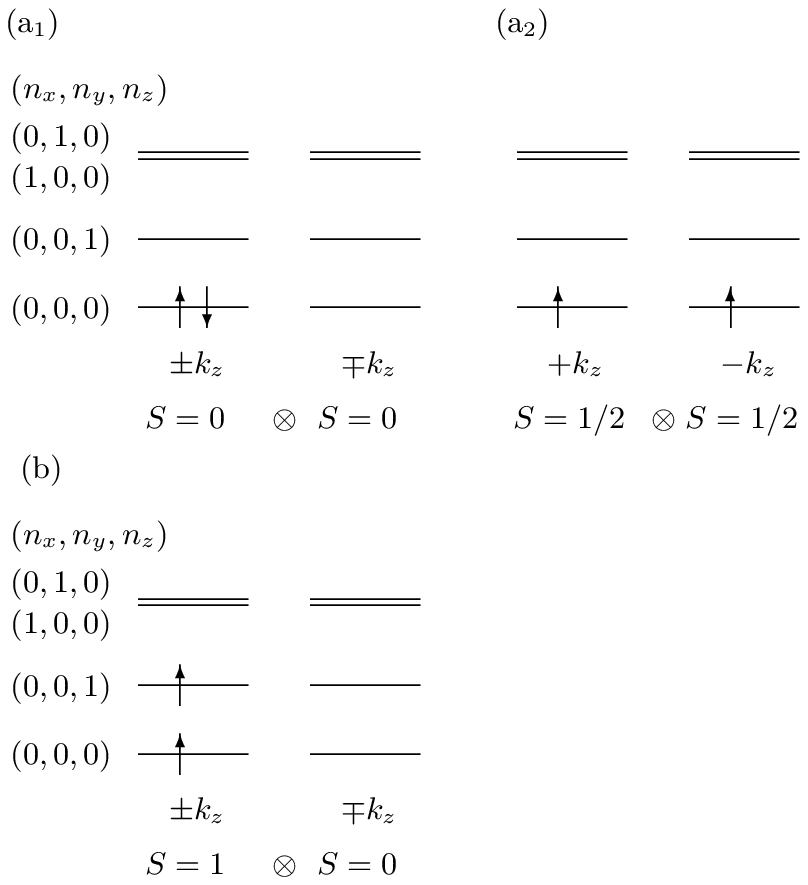}
\caption{Electronic configurations for the ground state with $N=2$
in isotropic quantum dots. $(n_x,n_y,n_z)$ are the orbital
quantum numbers for each one-electron level, whereas
$\pm k_z$ are the valley indices.
Configuration (a$_1$) or (a$_2$) is realized when the dot size
is $l=5$ and $10$ nm. They are degenerate in energy in the absence
of magnetic field. The total spin is $S_{\rm tot}=0$ and
$S_{\rm tot}=1/2 \otimes 1/2 =0 \oplus 1$, respectively.
Configuration (b) with $S_{\rm tot}=1$ is realized for $l=15$ nm.
\label{fig:N=2}}

\end{figure}

We begin with the one-electron energy levels. As shown in Eq.\
(\ref{eq:level}), the one-electron levels are labeled by
valley index and quantum numbers $(n_x,n_y,n_z)$ for the orbital motion.
The lowest levels belong to the valleys of $+k_z$ or $-k_z$
with $(n_x,n_y,n_z)=(0,0,0)$ (see horizontal lines in Fig.\ \ref{fig:N=2}).
The next levels are $(\pm k_z; 0,0,1)$, which are lower than
levels $(\pm k_z; 1,0,0)$ and $(\pm k_z; 0,1,0)$
in spite of the isotropic confinement. This is due to the
anisotropy of the effective mass
[$\varepsilon(\pm k_z; 1,0,0)-\varepsilon(\pm k_z; 0,0,1)=
\hbar (\sqrt{K / m_t^*}-\sqrt{K / m_l^*}) $].
We find that all the electrons are accommodated in valleys of $\pm k_z$
for $N=1$ to $5$ in all the cases of the dot size.

Regarding the electron-electron interaction between different valleys,
we obtain two significant results. (i) The Coulomb interaction between
$(\pm k_z; n_x,n_y,n_z)$ and $(\pm k_z; n_x',n_y',n_z')$ is the same as
that between $(\pm k_z; n_x,n_y,n_z)$ and $(\mp k_z; n_x',n_y',n_z')$.
This is because the envelope functions are equivalent for the valleys of
$+k_z$ and $-k_z$ in Eq.\ (\ref{eq:Coulomb}).
(ii) The exchange interaction is negligibly small between electrons in
different valleys, owing to the integration of rapidly oscillating factors
in Eq.\ (\ref{eq:exchange}). Therefore, there is no spin coupling
(Hund's coupling) between different valleys.

When the dot size is $l=5$ nm or $10$ nm, two electrons occupy the
lowest level in different valleys in the absence of Hund's coupling.
In consequence, the energies of different spin states are degenerate.
The configuration in Fig.\ \ref{fig:N=2}(a$_1$) has
the total spin of $S_{\rm tot}=0$,
whereas that in Fig.\ \ref{fig:N=2}(a$_2$)
has $S_{\rm tot}=1/2 \otimes 1/2 =0 \oplus 1$.
These configurations have the same energy.
In a small magnetic field, the Zeeman effect,
Eq.\ (\ref{eq:Zeeman}), makes the largest spin of
$S_{\rm tot}=1$ with $S_{{\rm tot},z}=1$.

When the dot size is $l=15$ nm, the ground state with $N=2$ has
another configuration;
one electron occupies an upper level to gain the exchange energy with
the other electron with parallel spin in a valley [Fig.\ \ref{fig:N=2}(b)].
Note that the exchange interaction, Eq.\ (\ref{eq:exchange}),
works effectively between electrons only in the same valley.
In general, the electronic states are determined
by the competition between
the electron-electron interaction and spacing of one-electron
energy levels. When the dot size is $l$, the former is
characterized by $E_{\rm int}=e^2/(4\pi\varepsilon l)$, whereas
the latter is by $E_{\rm levels}=\hbar^2/(m^*_t l^2)$.
The ratio of these energies, $E_{\rm int}/E_{\rm levels}=
m^*_t e^2 l/(4\pi\varepsilon \hbar^2)$, is $1.52$, $3.04$,
and $4.56$ for $l=5$ nm, $10$ nm, and $15$ nm, respectively.
When the ratio is larger, more electrons are accommodated in a valley
with parallel spins to gain the exchange energy.

\begin{figure}

\includegraphics[scale=1.00]{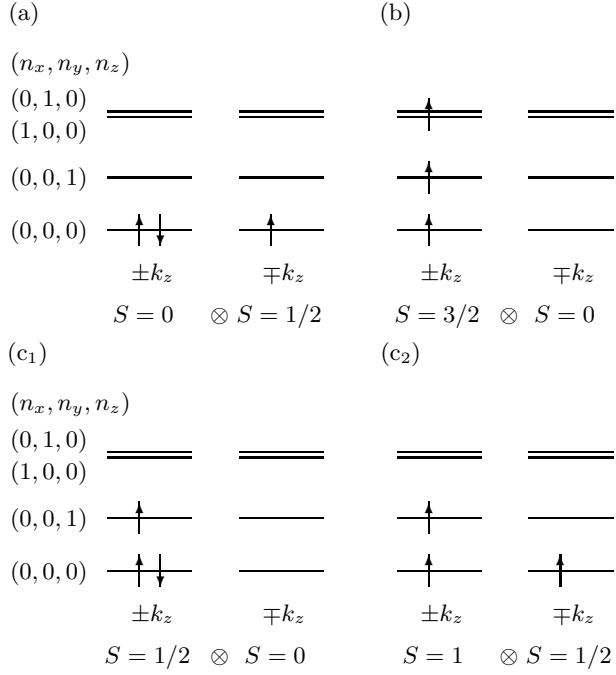}

\caption{Electronic configurations for the ground state with $N=3$
in isotropic quantum dots.
$(n_x,n_y,n_z)$ are the orbital quantum numbers for each one-electron
level, whereas $\pm k_z$ are the valley indices.
Configurations (a) and (b) are realized when the dot size is
$l=5$ nm and $15$ nm, respectively. Configuration (c$_1$) or (c$_2$) appears
for $l=10$ nm.
When the dot size is $l=$7.5 nm, configuration (a) appears at $B<9.47$ T,
whereas (c$_2$) appears $B>9.47$ T.
\label{fig:config3}}

\end{figure}

The ground state with $N=3$, shown in Fig.\ \ref{fig:config3}, can
be understood in a similar way. For the smallest dot ($l=5$ nm),
three electrons occupy the lowest levels [$S_{\rm tot}=1/2$;
Fig.\ \ref{fig:config3}(a)].
For the largest dot ($l=15$ nm), they occupy upper levels
in a valley, making $S_{\rm tot}=3/2$, to gain the exchange energy
[Fig.\ \ref{fig:config3}(b)].
For the intermediate case ($l=10$ nm), two electrons make
a spin triplet, occupying two levels in a valley,
while an electron is accommodated in the lowest level
in the same valley (c$_1$) or in the other equivalent valley (c$_2$).
These configurations have the same energy but different spin,
(c$_1$) $S_{\rm tot}=1/2$ and
(c$_2$) $S_{\rm tot}=1 \otimes 1/2 =1/2 \oplus 3/2$, the largest spin
of which is realized in a magnetic field.

With increasing magnetic field, the Zeeman effect, Eq.\ (\ref{eq:Zeeman}),
makes larger spin states more favorable. For $N=3$, the ground state could
change from configuration
Fig.\ \ref{fig:config3}(a) ($S_{\rm tot}=1/2$) at $B<B_{\rm c}$
to configuration Fig.\ \ref{fig:config3}(b) ($S_{\rm tot}=3/2$) at
$B>B_{\rm c}$. Such a transition is observed at $B_{\rm c}=9.47$ T when
the dot size is $l=7.5$ nm. (The same transition takes place at much
larger $B_{\rm c}$ when $l=5$ nm.) We find a similar transition with
$N=4$, from $S_{\rm tot}=1$ to $S_{\rm tot}=2$ at $B_{\rm c}=3.44$ T,
when the dot size is $l=7.5$ nm.

\subsection{Addition energies}

For the comparison with the experimental results, we calculate addition
energies which correspond to peak positions of the Coulomb
oscillation.\cite{Ta,Kouwenhoven}
The addition energy of the $N$th electron on the dot is given by
\begin{eqnarray}
\mu_N = E_N - E_{N-1},
\end{eqnarray}
where $E_N$ is the ground-state energy with $N$ electrons.
Figure \ref{fig:addition} shows $\mu_N$
($N=1$ to 5), as functions of magnetic field $B$.
The slope of $\mu_N(B)$ is given by
\[
-g^* \mu_{\rm B} [S_{\rm tot}(N)-S_{\rm tot}(N-1)],
\]
which reflects the total spin $S_{\rm tot}(N)$ with $N$ electrons.

In Fig.\ \ref{fig:addition}(a), the dot size is the smallest ($l=5$ nm).
The one-electron levels are consecutively occupied with increasing the
number of electrons $N$.
For $N=2$, the electrons are accommodated in different valleys with
parallel spins in a magnetic field,
as discussed previously. In consequence,
$S_{\rm tot}(N)=1/2$, $1$, $1/2$, $0$,
and $1/2$ for $N=1$ to $5$, respectively.
This results in the slope of $- g^* \mu_{\rm B}/2$
for $\mu_1(B)$, $\mu_2(B)$, $\mu_5(B)$, and
$+ g^* \mu_{\rm B}/2$ for $\mu_3(B)$, $\mu_4(B)$.

In Fig.\ \ref{fig:addition}(b), the dot is the largest ($l=15$ nm).
To gain the exchange energy, the electrons occupy
upper levels in a valley, making the highest spin, $S_{\rm tot}(N)=N/2$.
Accordingly, $\mu_N(B)$ ($N=1$ to $5$) are parallel to each other,
with the same slope of $-g^* \mu_{\rm B}/2$.
When the dot size is $l=10$ nm,
the highest spin of $S_{\rm tot}(N)=N/2$ is also realized.
The slopes of $\mu_N(B)$ ($N=1$ to $5$) are the same as those in
Fig.\ \ref{fig:addition}(b).

In Fig.\ \ref{fig:addition}(c), the dot size is $l=7.5$ nm.
$\mu_{3}(B)$ shows a kink at $B_{\rm c}=9.47$ T, which is due to
the transition of the ground state with $N=3$ from $S_{\rm tot}(3)=1/2$ to
$3/2$. Since $S_{\rm tot}(2)=1$, the slope of $\mu_{3}(B)$
is $g^* \mu_{\rm B}/2$ and $-g^* \mu_{\rm B}/2$
at $B<B_{\rm c}$ and $B>B_{\rm c}$, respectively.
This transition results in a kink of $\mu_{4}(B)$ at the same
magnetic field [with $S_{\rm tot}(4)=2$, the slope of $\mu_{4}(B)$
is $-3g^* \mu_{\rm B}/2$ at $B<B_{\rm c}$, $-g^* \mu_{\rm B}/2$
at $B>B_{\rm c}$].
Similarly, the transition of the ground state with $N=4$ makes a pair of
kinks on $\mu_{4}(B)$ and $\mu_{5}(B)$ at $B_{\rm c}=3.44$ T.
Such pairs of kinks of adjacent addition energies have been observed
in the Coulomb oscillation.\cite{Ro1}

At $3.44$ T$<B<9.47$ T in Fig.\ \ref{fig:addition}(c),
$\mu_{4}(B)$ has a slope of
$-3g^* \mu_{\rm B}/2$ (indicated by dotted line).
Then the conductance through the quantum dot
should be suppressed since the addition of the fourth electron is
forbidden by the spin selection rule: $S_{\rm tot}=1/2$ with $N=3$ and
$S_{\rm tot}=2$ with $N=4$. This is called spin blockade.\cite{Weinmann}
The spin blockade in Si quantum dots has been found experimentally.\cite{Ro1}

\section{Anisotropic confinemnet}
\label{sec:shape}

In the previous section, we have examined the isotropic confinement
to illustrate the basic properties of Si quantum dots. In this section,
we investigate an anisotropic confinement.
Quantum dots, which are fabricated in the middle of Si wires
with oxidation technique,\cite{Tak,Ho,Hi,Ro1,Ro2,On} may be
elongated along the wires. We adopt an elliptical potential,
elongated in [110] direction of the lattice,\cite{Ho}
\begin{equation}
V(x,y,z)=\frac{1}{2} \left[ K' \left( \frac{x+y}{\sqrt{2}} \right)^2
+ K \left( \frac{-x+y}{\sqrt{2}} \right)^2 + K z^2 \right].
\end{equation}
with $K'<K$. By the transformation of the axes to $X=(x+y)/\sqrt{2}$,
$Y=(-x+y)/\sqrt{2}$, and $z$, it is rewritten as
$V(X,Y,z)=(1/2)( K' X^2 + K Y^2 + K z^2 )$.
For valleys of $\pm k_z$, the one-electron energy levels are given by
\begin{eqnarray}
\lefteqn{\varepsilon(\pm k_z; n_X,n_Y,n_z)}
\nonumber \\
&=& \hbar \sqrt{\frac{K'}{m^*_t}}(n_X +\frac{1}{2})
+ \hbar \sqrt{\frac{K}{m^*_t}}(n_Y +\frac{1}{2})
+ \hbar \sqrt{\frac{K}{m^*_l}}(n_z +\frac{1}{2}).
\nonumber \\
\
\label{eq:level2}
\end{eqnarray}
We change the size of the confinement potential in $[110]$ direction,
$l_X=2\sqrt{\hbar/m^*_t \omega_{Xt}}$
where $\omega_{Xt}=\sqrt{K'/m^*_t}$, as 5 nm, 10 nm, and 15 nm.
The size in $[\bar{1}10]$ and $[001]$ directions is fixed at
$l_Y=l_z=5$ nm, where
$l_Y=2\sqrt{\hbar/m^*_t \omega_{Yt}}$,
and
$l_z=2\sqrt{\hbar/m^*_t \omega_{zt}}$
with $\omega_{Yt}=\omega_{zt}=\sqrt{K/m^*_t}$.

\begin{figure}

\includegraphics[scale=1.00]{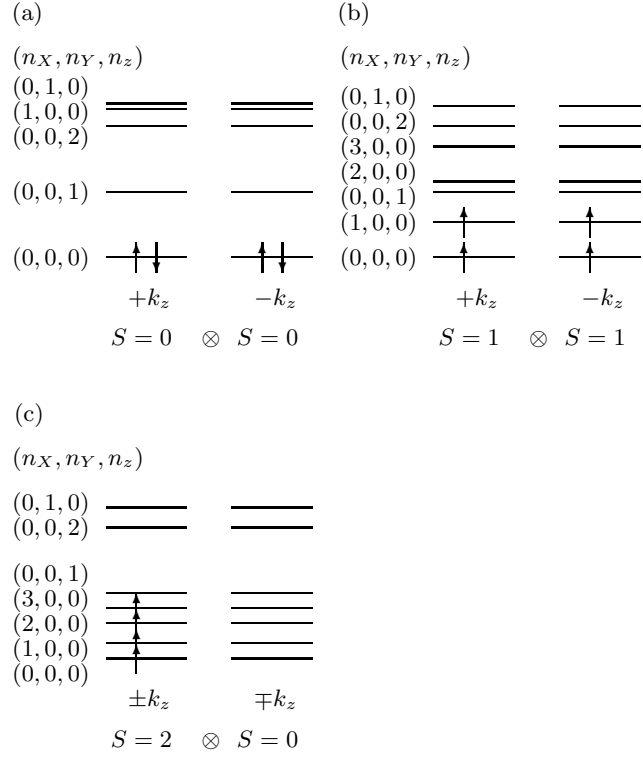}
\caption{Electronic configurations for the ground state with $N=4$
in elliptical quantum dots.
$(n_X,n_Y,n_z)$ are the orbital quantum numbers for each one-electron
level, whereas $\pm k_z$ are the valley indices.
The size of the dot in [110] direction is (a) $l_X=5$ nm,
(b) 10 nm, and (c) 15 nm,
whereas the size in the other directions is fixed at $l_Y=l_z=5$ nm.
\label{fig:shape-dep}}

\end{figure}

Figure \ref{fig:shape-dep} shows the electronic configuration of the
ground state with $N=4$ when (a) $l_X=5$ nm (isotropic dot),
(b) $l_X=10$ nm, and (c) $15$ nm.
As the shape of the dot is more elongated in [110] direction,
the energy levels $(n_X,0,0)$ ($n_X=1,2,3\ldots$) are lowered
[see Eq.\ (\ref{eq:level2}) with smaller $K'$].
Electrons begin to occupy these levels: (a) $(\pm k_z; 0,0,0)$ only,
(b) $(\pm k_z; 0,0,0)$ and
$(\pm k_z; 1,0,0)$, and (c) $(0,0,0)$ to $(3,0,0)$ in a valley.
This means that the electrons tend to be populated in the elongated
direction. Reflecting small level spacings between $(n_X,0,0)$,
high-spin states are realized more often:
(a) $S_{\rm tot}=0$, (b) $S_{\rm tot}=1 \otimes 1=0 \oplus 1 \oplus 2$,
and (c) $S_{\rm tot}=2$.
We find that the ground state has the largest spin of $S_{\rm tot}(N)=N/2$
for $N=1$ to $5$ in elongated dots with $l_X=10$ nm and $15$ nm
($l_Y=l_z=5$ nm) in a magnetic field.
The magnetic-field dependence of addition energies
$\mu_N$ shows the same slope of $-g^* \mu_{\rm B}/2$, which is
similar to the situation in Fig.\ \ref{fig:addition}(b).

\section{Intervalley Scattering}
\label{sec:intervalley}

In the present section, we discuss the effects of intervalley scattering.
The intervalley scattering is not relevant when the size of the dots is
much larger than the lattice constant and confinement potential is
sufficiently smooth. However, the scattering by impurities within the dots,
sharp edges of the dots, etc., mixes the electronic states
in different valleys. As an example, let us consider an impurity potential
at $\bm{R}$ inside a quantum dot,
\[
V_{\rm imp}(\bm{r})=V_0 \delta(\bm{r}-\bm{R}),
\]
to the lowest order of the perturbation.
The degenerate levels of $(\pm k_z; n_x,n_y,n_z)$
are split into two by this potential,
\[
\varepsilon(a; n_x,n_y,n_z) = \varepsilon(\pm k_z; n_x,n_y,n_z),
\]
and
\[
\varepsilon(b; n_x,n_y,n_z) = \varepsilon(\pm k_z; n_x,n_y,n_z)
+ \Delta_{n_x,n_y,n_z},
\]
where $\Delta_{n_x,n_y,n_z}=2V_0 |F_{\pm k_z}^{n_x,n_y,n_z}(\bm{R})|^2
|u_{\pm k_z}(\bm{R})|^2$, as shown in Fig.\ \ref{fig:impurity}(a).
The corresponding wavefunctions are expressed as
\begin{eqnarray*}
\psi_{a}^{n_x,n_y,n_z}(\bm{r}) &=& \frac{1}{\sqrt{2}}
\{ \psi_{+k_z}^{n_x,n_y,n_z}(\bm{r})
-e^{-i \theta}\psi_{-k_z}^{n_x,n_y,n_z}(\bm{r}) \}, \\
\psi_{b}^{n_x,n_y,n_z}(\bm{r}) &=& \frac{1}{\sqrt{2}}
\{ \psi_{+k_z}^{n_x,n_y,n_z}(\bm{r})
+e^{-i \theta}\psi_{-k_z}^{n_x,n_y,n_z}(\bm{r}) \},
\end{eqnarray*}
with $\theta={\rm arg} \langle +k_z;n_x,n_y,n_z | V_{\rm imp} |
-k_z;n_x,n_y,n_z \rangle$.
For the electron-electron interactions, we find that
(i) the Coulomb interaction between
$(a; n_x,n_y,n_z)$ and $(a; n_x',n_y',n_z')$ is identical to
that between $(b; n_x,n_y,n_z)$ and $(b; n_x',n_y',n_z')$, and also to
that between $(a; n_x,n_y,n_z)$ and $(b; n_x',n_y',n_z')$.
(ii) The exchange interaction is negligibly small between
$(a; n_x,n_y,n_z)$ and $(b; n_x',n_y',n_z')$.

The intervalley scattering influences the spin states in a
spherical dot of $l=5$ nm. In the absence of intervalley scattering,
the first two electrons occupy the degenerate levels of $(\pm k_z; 0,0,0)$
with parallel spins [$S_{\rm tot}(2)=1$] in a small magnetic field
[Fig.\ \ref{fig:N=2}(a$_2$)]. In the presence of intervalley
scattering, the two electrons occupy the lowest level of $(a; 0,0,0)$
with anti-parallel spins [$S_{\rm tot}(2)=0$], as shown in
Fig.\ \ref{fig:impurity}(a). For $N=1$ to 5, the spin states
become $S_{\rm tot}(N)=1/2$, $0$, $1/2$, $0$, and $1/2$, respectively.

The addition energies are shown in Fig.\ \ref{fig:impurity}(b),
as functions of magnetic field, in the case of
$\Delta_{0,0,0}=2$ meV. The slopes of $\mu_N$ are
$-g^* \mu_{\rm B}/2$
or $+g^* \mu_{\rm B}/2$ alternately with increasing $N$,
reflecting the above-mentioned spin states.
In the case of $\Delta_{0,0,0}=0.4$ meV, the transition of the
ground state with $N=2$ is observed at the magnetic field of
$B_{\rm c}=3.45$ T, where $g^* \mu_{\rm B}B_{\rm c}=\Delta_{0,0,0}$.
For $B<B_{\rm c}$, two electrons occupy
level $(a; 0,0,0)$, making $S_{\rm tot}(2)=0$, as shown in Fig.\
\ref{fig:impurity}(a).
For $B>B_{\rm c}$, one electron occupies level $(a; 0,0,0)$ and the other
occupies level $(b; 0,0,0)$ with parallel spins [$S_{\rm tot}(2)=1$],
to gain the Zeeman energy. This transition makes a pair of kinks
on $\mu_2$ and $\mu_3$ [see Fig.\ \ref{fig:impurity}(c)].

When the dot size is much larger than 10 nm, the intervalley scattering
never changes the spin states.
Electrons occupy the levels of $(a; 0,0,0)$,
$(a; 0,0,1)$, $(a; 1,0,0)$, $\cdots$, consecutively with parallel
spins to gain the exchange energy
[not the levels of $(b; n_x,n_y,n_z)$ owing to the absence of
exchange interactions between $(b; n_x,n_y,n_z)$ and $(a; n_x',n_y',n_z')$].
Then the largest spins of $S_{\rm tot}(N)=N/2$ are realized.
The addition energies $\mu_N(B)$ ($N=1$ to $5$)
have the same slope of $-g^* \mu_{\rm B}/2$, in the same way as
in Fig.\ \ref{fig:addition}(b).

\section{Conclusions and Discussion}
\label{sec:conclusion}

We have theoretically examined the electronic states in Si dots,
multivalley artificial atoms. (i) In spherical quantum dots smaller than
$\sim 10$ nm, the one-electron levels in equivalent valleys
are occupied consecutively with increasing the number of electrons $N$.
High-spin states are attributable to the absence of spin couplings
between different valleys. This is quite different from the case of
GaAs quantum dots, where high spins appear due to the exchange interaction
in exceptional cases with orbital degeneracy (Hund's rule).
(ii) In quantum dots much larger than $10$ nm, the electrons tend to occupy
upper levels in a valley, making the largest spin of $S_{\rm tot}=N/2$,
to gain the exchange energy. (iii) In elliptical quantum dots,
the electrons tend to be populated in the elongated direction.
High-spin states often appear.
(iv) In the presence of intervalley scattering,
the degenerate energy levels in different valleys are split.
In small dots, this results in the lowest spin states
($S_{\rm tot}=0$ or $1/2$).
The spin states in large dots, $S_{\rm tot}=N/2$, are not influenced by
the intervalley scattering.

In the present study, the electronic states are represented by
single configurations of the occupation of
one-electron levels by $N$ electrons, for
semi-quantitative discussion. We have neglected the configuration interaction,
or correlation effect.
(It should be noted that the configuration interaction vanishes between
electrons in different valleys for the same reason as the exchange interaction
vanishes.) In this approximation, the energy of
low-spin states is overestimated compared with the energy of high-spin
states. Hence we should have underestimated the dot size where the
highest spins appear.
The correlation effect becomes more important as the dot size is larger
($E_{\rm int}/E_{\rm levels}$ is larger) for quantitative
discussion.\cite{Eto3} The consideration of the correlation effect with
multivalley structure in Si dots requires further study.

\begin{figure}

\includegraphics[scale=1.00]{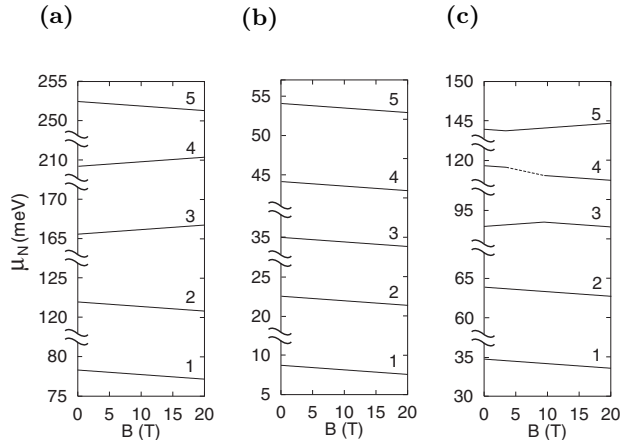}

\caption{Addition energies, $\mu_N$ ($N=1$ to $5$),
as functions of magnetic field $B$, in isotropic quantum dots. The dot size is
(a) $l=5$ nm, (b) $15$ nm, and (c) $7.5$ nm.
In (c), the spin blockade occurs for $\mu_4(B)$ at $3.44$ T$<B<9.47$ T
(dotted line).
\label{fig:addition}}

\end{figure}

\begin{figure}

\includegraphics[scale=1.00]{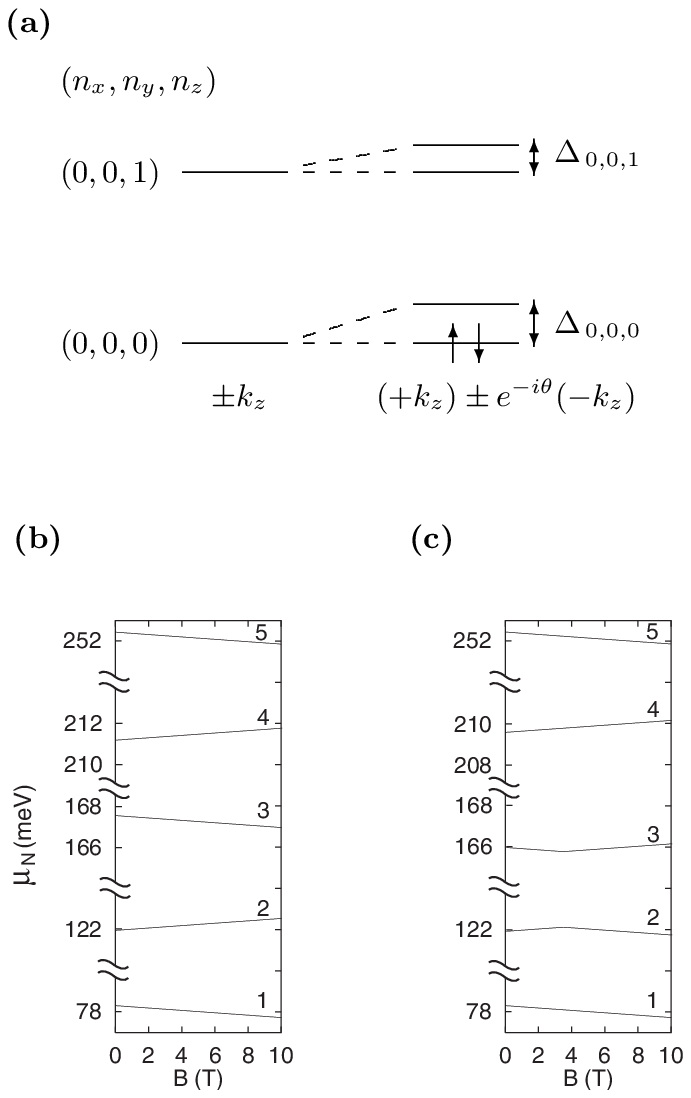}

\caption{(a) The effect of intervalley scattering on the one-electron
energy levels. The degenerate levels of $(\pm k_z;n_x,n_y,n_z)$ are
split into two.
(b), (c) Addition energies, $\mu_N$ ($N=1$ to $5$),
as functions of magnetic field $B$, in isotropic quantum dots of $l=5$ nm.
The energy splitting by the intervalley scattering is
(b) $\Delta_{0,0,0}=2$ meV and (c) $\Delta_{0,0,0}=0.4$ meV.
\label{fig:impurity}}

\end{figure}

The spin states in Si quantum dots depend on both their size and shape.
Our calculated results cannot be compared directly with experimental
results of the Coulomb oscillation\cite{Ro1,Ro2,On} since the exact shape of
quantum dots is not known. Roughly speaking, the magnetic-field dependence
of peak positions in the observed Coulomb oscillation is similar to that in
Figs.\ \ref{fig:impurity}(b), (c) rather than that in
Figs.\ \ref{fig:addition}(a)-(c), which might imply that low-spin states
are realized in small dots and in the presence of intervalley scattering.
Qualitatively,
pairs of kinks have been observed,\cite{Ro1} which indicate the
transitions of the ground state. The spin blockade with
$|S_{\rm tot}(N+1)-S_{\rm tot}(N)| \ge 3/2$ has also been reported.\cite{Ro1}
We have shown that the spin states can be changed by the intervalley
scattering. If the strength of the intervalley scattering were tuned,
e.g., using STM tips, the spin states in Si dots could be controlled.

Recently, quantum computing devices have been proposed, utilizing electron
spins in quantum dots.\cite{Loss} For the devices, the decoherence time of
electron spins must be much longer than the gate operation time.
Main origins of the decoherence are spin-orbit interaction and hyperfine
interaction with nuclear spins in the dots. In Si, the spin-orbit interaction
is much weaker than in GaAs. The hyperfine interaction is also weaker due to
smaller isotope ratio with nuclear spins ($^{29}$Si of 4.7\% only).
Hence a longer decoherence time is expected in Si quantum dots than in
GaAs quantum dots. Therefore, Si dots may be advantageous to the quantum
computing devices. In Si dots, high-spin states are realized easily,
which could be utilized for the devices.

\begin{acknowledgments}
This work was partially supported by a Grant-in-Aid for
Scientific Research in Priority Areas ``Semiconductor Nanospintronics"
(No.\ 14076216) of The Ministry of Education, Culture, Sports, Science
and Technology, Japan.
\end{acknowledgments}

\newpage

\begin{thebibliography}{99}

\bibitem{Ta} S.\ Tarucha, D.\ G.\ Austing, T.\ Honda,
R.\ J.\ van der Hage, and L.\ P.\ Kouwenhoven,
Phys.\ Rev.\ Lett.\ {\bf 77}, 3613 (1996).

\bibitem{Kouwenhoven} L.\ P.\ Kouwenhoven, T.\ H.\ Oosterkamp,
M.\ W.\ S.\ Danoesastro, M.\ Eto, D.\ G.\ Austing, T.\ Honda, and S.\ Tarucha,
Science {\bf 278}, 1788 (1997).

\bibitem{qd1}
P.\ A.\ Maksym and T.\ Chakraborty, Phys.\ Rev.\ Lett.\
{\bf 65}, 108 (1990).

\bibitem{qd2}
M.\ Wagner, U.\ Merkt, and A.\ V.\ Chaplik, Phys.\ Rev.\ B {\bf 45}, 1951
(1992).

\bibitem{qd3}
P.\ Hawrylak, Phys.\ Rev.\ Lett.\ {\bf 71}, 3347 (1993).

\bibitem{qd4}
D.\ Pfannkuche and S.\ E.\ Ulloa,
Phys.\ Rev.\ Lett.\ {\bf 74}, 1194 (1995).

\bibitem{Eto1}
M.\ Eto, Jpn.\ J.\ Appl.\ Phys.\ {\bf 36}, 3924 (1997).

\bibitem{Eto2}
M.\ Eto, Jpn.\ J.\ Appl.\ Phys.\ {\bf 38}, 376 (1999).

\bibitem{Eto3}
M.\ Eto, Jpn.\ J.\ Appl.\ Phys.\ {\bf 40}, 1929 (2001),
and related references cited therein.

\bibitem{Tak} Y.\ Takahashi, M.\ Nagase, H.\ Namatsu, K.\ Kurihara,
K.\ Iwadate, Y.\ Nakajima, S.\ Horiguchi, K.\ Murase,
and M.\ Tabe,
Electron. Lett.\ {\bf 31}, 136 (1995).

\bibitem{Ho} S.\ Horiguchi, M.\ Nagase, K.\ Shiraishi, H.\ Kageshima,
Y.\ Takahashi, and K.\ Murase,
Jpn.\ J.\ Appl.\ Phys.\ {\bf 40}, L29 (2001),
and related references cited therein.

\bibitem{Hi}
T.\ Hiramoto, H.\ Ishikuro, T.\ Fujii, G.\ Hashiguchi, and T.\ Ikoma,
Jpn.\ J.\ Appl.\ Phys.\ {\bf 36}, 4139 (1997).

\bibitem{Sa} M.\ Saitoh, N.\ Takahashi, H.\ Ishikuro, and T.\ Hiramoto,
Jpn.\ J.\ Appl.\ Phys.\ {\bf 40}, 2010 (2001).

\bibitem{Ro1} L.\ P.\ Rokhinson, L.\ J.\ Guo, S.\ Y.\ Chou, and D.\ C.\ Tsui,
Phys.\ Rev.\ B {\bf 63}, 35321 (2001).

\bibitem{Ro2} L.\ P.\ Rokhinson, L.\ J.\ Guo, S.\ Y.\ Chou, and D.\ C.\ Tsui,
Phys.\ Rev.\ Lett.\ {\bf 87}, 166802 (2001).

\bibitem{On} K.\ Ono, private communications.

\bibitem{Na} The multivalley structure has been taken into account by
K.\ Natori, T.\ Uehara, and N.\ Sano, in the transport properties of
single-electron transistors using Si quantum dots
[Jpn.\ J.\ Appl.\ Phys.\ {\bf 39}, 2550 (2000)]. They have examined
a charging model for electron-electron interaction, and hence disregarded
spin states in Si quantum dots.

\bibitem{Ro3} L.\ P.\ Rokhinson, L.\ J.\ Guo, S.\ Y.\ Chou, and D.\ C.\ Tsui,
Phys.\ Rev.\ B {\bf 60}, R16319 (1999).

\bibitem{Loss}
D.\ Loss and D.\ P.\ DiVincenzo, Phys.\ Rev.\ A {\bf 57}, 120 (1998).

\bibitem{Lu} J.\ M.\ Luttinger and W.\ Kohn,
Phys.\ Rev.\ {\bf 97}, 869 (1955).

\bibitem{Shi} K.\ Shiraishi, M.\ Nagase, S.\ Horiguchi, H.\ Kageshima,
M.\ Uematsu, Y.\ Takahashi, and K.\ Murase,
Physica E {\bf 7}, 337 (2000).

\bibitem{orbital} In Si,
even the lighter effective mass of $m^*_t=0.19\,m_0$ is
much larger than the effective mass $m^*=0.067\,m_0$ in GaAs.
Hence the orbital effect is smaller than in
GaAs quantum dots.\cite{Kouwenhoven,Eto3}

\bibitem{Weinmann}
D.\ Weinmann, W.\ H\"{a}usler, and B.\ Kramer,
Phys.\ Rev.\ Lett.\ {\bf 74}, 984 (1995).

\end{thebibliography}
\end{document}